\documentclass[10pt,conference,a4paper]{IEEEtran}
\usepackage{graphicx}
\usepackage{epstopdf}
\usepackage[usenames,dvipsnames]{xcolor}
\usepackage{setspace}
\usepackage{cite}
\usepackage[hidelinks]{hyperref}
\newtheorem{theorem}{Theorem}
\usepackage{soul}
\usepackage{booktabs}
\usepackage{colortbl}

\begin{document}
\title{A Policy Switching Approach to Consolidating Load Shedding and Islanding Protection Schemes}
\author{
\IEEEauthorblockN{Rich Meier, Eduardo Cotilla-Sanchez, Alan Fern}
\IEEEauthorblockA{School of Electrical Engineering \& Computer Science\\ Oregon State University, Corvallis, OR 97331 USA\\Email: \{meierr,ecs,afern\}@eecs.oregonstate.edu}
}
\maketitle


\begin{abstract}
\label{abstract}
\textbf{In recent years there have been many improvements in the reliability of critical infrastructure systems. Despite these improvements, the power systems industry has seen relatively small advances in this regard. For instance, power quality deficiencies, a high number of localized contingencies, and large cascading outages are still too widespread. Though progress has been made in improving generation, transmission, and distribution infrastructure, remedial action schemes (RAS) remain non-standardized and are often not uniformly implemented across different utilities, independent system operators (ISOs), and regional transmission organizations (RTOs). Traditionally, load shedding and islanding have been successful protection measures in restraining propagation of contingencies and large cascading outages. This paper proposes a novel, algorithmic approach to selecting RAS policies to optimize the operation of the power network during and after a contingency. Specifically, we use policy-switching to consolidate traditional load shedding and islanding schemes. In order to model and simulate the functionality of the proposed power systems protection algorithm, we conduct Monte-Carlo, time-domain simulations using Siemens PSS/E. The algorithm is tested via experiments on the IEEE-39 topology to demonstrate that the proposed approach achieves optimal power system performance during emergency situations, given a specific set of RAS policies. 
}

\end{abstract}


\section{Introduction}
\label{intro}
Power systems contingencies continue to have negative impacts, not only on the ability to provide immediate electrical power where needed, but more significantly on the long-term sustainability of our economic, societal, and political institutions that rely on a dependable source of energy. Power systems engineers and scientists are continually striving to find methods that will mitigate the effects of both exogenous and endogenous system perturbations. Extreme variability in load or generation, machine or component failure, adverse weather, and even acts of terrorism \cite{Paper026} are some examples of the former. Operator error, improper device settings or maintenance, and miscalculation of generation or load dispatch are a few simple examples of the latter. While in both cases it is important to prevent entering into fragile situations, it is often the case, in practice, that remedial action schemes (RAS\footnote{They are also known as Special Protection Schemes (SPS) or System Integrity Protection Schemes (SIPS).}) must be implemented during a contingency to avoid cascading outages. Thus, designing power system protection methods that are robust, fair, efficient, and reliable is crucial. 

The breadth of RAS policies is vast and continually expanding. Traditionally, two protection actions have proven to be highly effective as part of an RAS - namely load shedding and islanding. Load shedding has many permutations ranging from homogeneous, naive techniques \cite{Paper001,Paper013} to highly complex and adaptive \cite{Paper012,Paper009,Paper005}. Load shedding is a powerful scheme because it fundamentally seeks to optimize the power flow so that load and generation match precisely. Also, depending on its complexity, the computation time is often small enough to merit on-line operation.  Islanding, on the other hand, is a relatively new method for protection. Its main purpose is to isolate functional sections of the power system so that a contingency does not cascade and become more severe. Some studies have been done with regard to islanding's effectiveness on both micro and macro scale grid topologies \cite{Paper021} \cite{Paper022}. Others have started to consider the effects of having distributed generation (DG) sources contained within islands \cite{Paper019}. Despite the many factors that influence the use of islanding, it is largely agreed upon that this protection method is useful for improving the operation of a power system in some contingency situations \cite{Paper023}. Typically islanding policies have been determined offline by simulating a large set of different contingencies and finding sections of the grid that can continue to operate independently of adjacent grid infrastructure. Reference \cite{Paper023} also proposes one of the first on-line islanding algorithms.

Given the variety of load shedding and islanding approaches, along with their various combinations, the number of possible RAS policies (just considering these two schemes) is vast. Unfortunately, during emergency situations it is difficult to determine which such policy will optimally improve network operation after a contingency. The main contribution of this paper is to provide a novel approach to selecting such RAS policies in a manner that maintains strong performance guarantees. Specifically, we introduce the framework of \textit{policy switching}. Policy switching is an algorithmic approach to selecting and executing the policy that exhibits the best performance according to Monte-Carlo simulation techniques \cite{Paper024}. As time progresses, this approach may switch between multiple policies, depending on the specific way that the grid system evolves. This methodology has been used in machine learning and artificial intelligence spheres to improve the outcomes of multiple control problems ranging from real-time strategy (RTS) games to computer networking. Grid protections provide similar control issues in that there is a wide breadth of RAS policies to choose from. Ideally, the optimal solution will be found in an automated and real-time manner. 

The remainder of this paper will first address, in more detail, the RAS policies of load shedding and islanding. Subsequently, further background on the theory of policy switching (Markov Decision Processes) and its potential applications in power systems will be discussed (Sections \ref{ras}, \ref{mdp} and \ref{ps} respectively). Section \ref{methodology} will discuss the scope and the notable assumptions made in the simulations, as well as introduce the specific algorithm that is simulated on a model of IEEE-39. Following that, Section \ref{exp} will highlight the details of the experiments performed. The paper concludes with Section \ref{conclude} which explains key takeaways as well as the authors' ideas for algorithm improvements and future policy switching work in the field of power systems.


\section{Background}
\label{back}

\subsection{Remedial Action Schemes}
\label{ras}
This section reviews relevant load shedding and islanding approaches in order to lay the framework for why intelligent selection algorithms are useful.

\subsubsection{Load Shedding}
Load shedding is characterized by the ability to alter the amount of electrical power consumed by a specific load bus (or by a set of load buses). In some more advanced methodologies only particular loads are partially or fully shut down, whereas more naive methods decrease load power consumption at all of the loads in the network. Advanced approaches, often labeled as adaptive, are characterized by the ability to do one or more of the following: determine the severity of a contingency and shut down loads appropriately, determine where a contingency occurs and shut down loads spatially or electrically near the contingency, assign significance to particular loads that are of higher priority (i.e., hospitals, schools, data centers, etc.), as well as base decisions off of bus voltages, system frequency, and the time derivative of voltage and frequency \cite{Paper005,Paper008,Paper009,Paper012}.

In order to test the policy of load shedding, we select a simple approach that is commonly used in utility protective methods today.  That is, we select traditional, under-voltage load shedding. This RAS is specifically chosen because of its ease of implementation, as well as its aptness to be performed on-line. Traditional load shedding selects each operating, controllable load in the system and reduces its power consumption by a fixed ratio $R$. More formally:
\begin{equation}
\label{eqn1}
L_{\textnormal{final}} = \sum_{k=i}^{j}L_{k}*(1-R) 
\end{equation}
 
Where $L_{\textnormal{final}}$ is defined as the MVA power demand after load shedding, and $L_k$ is each operational load at bus $k$, where $k$ iterates over in-service load buses from the $i$th load bus to the $j$th load bus.

\subsubsection{Islanding}
Islanding techniques are relatively new to the power systems protection arsenal. One way of performing islanding analysis is to simulate a large set of possible contingencies (e.g., $N-1-1$, $N-2$) on the network topology in question. For each specific contingency, an islanding scheme is selected that best allows the system to be saved from severe off-nominal operation. An operator or an automated RAS can then recall a specific islanding scheme if ever the matching contingency occurs. Some new islanding methodologies that have been recently proposed that use slow-coherency, active and reactive power balancing, and distributed generation considerations \cite{Paper019,Paper021,Paper022,Paper023}.

\begin{figure}[!h]
\begin{center}
\includegraphics[scale=.38]{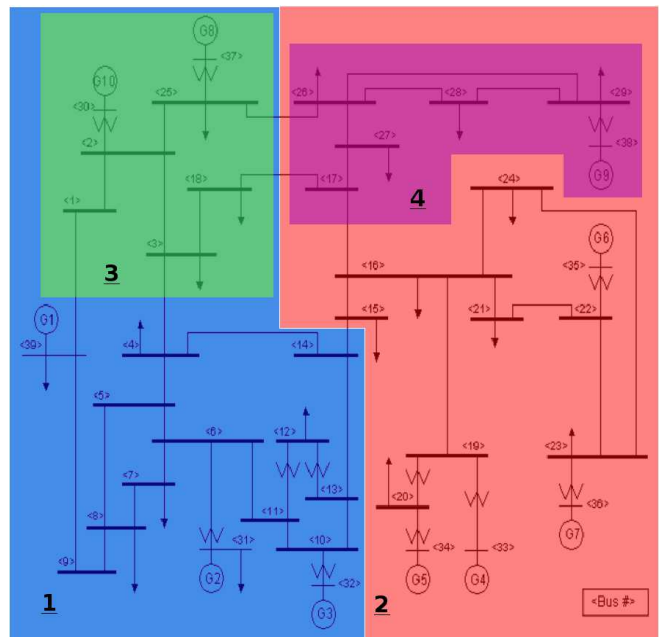}
\caption{\textit{The example islanding rules for the policy switching experiments.}}
\label{fig1}
\end{center}
\end{figure}

For the purpose of our experiments, we define two levels of islanding capability that have been determined offline. Similar to current methods used in grid operations, we have pre-determined where to island based on balancing load and generation in each island~\cite{Paper047}. Fig.~\ref{fig1} depicts the IEEE Case 39 topology with our example islanding scheme. The general concept of our islanding method is to split the case, electrically, in half. This illustrative rule in Fig.~\ref{fig1} can be generalized to a proper measure of electrical distance that yields relatively stable islands in a slow-coherency sense, as proposed in \cite{Cotilla-Sanchez:2013a}. Due to the small size of Case 39 we determined that more than two iterations of this policy created islands that were not autonomous and therefore ineffective. For the first iteration, we partition the case into \textit{Island 1} and \textit{Island 2} (labeled and shadowed in blue and red in Fig.~\ref{fig1}). The second level of islanding allows another split in half. Specifically, two subislands contained within \textit{Island 1} and \textit{Island 2} are created. As seen in Fig.~\ref{fig1}, the two new sub-partitions are labeled \textit{Island 3} and \textit{Island 4} (shadowed in green and purple).

\subsection{Markov Decision Processes}
\label{mdp}

We formalize the control problem of choosing amongst RAS policies within the framework of Markov Decision Processes (MDPs) \cite{puterman2009}. MDPs are a widely used mathematical model of controllable systems with stochastic dynamics. By modeling a control problem as an MDP, it is possible to draw on decades of research that has produced rich theory and computational solutions. Here we describe the necessary MDP concepts for this paper and in the next section describe our chosen computational approach of policy switching.

An MDP is a tuple $(S,A,P,R)$, where $S$ is a set of system states and $A$ is a set of possible control actions. For example, in a power system network, each state would correspond to a possible configuration of the joint state variables in the network. Each action might correspond to the control actions available to a network operator, e.g.~opening a transmission line. The third MDP component, $P$, is a conditional transition distribution defined such that $P(s' \;|\; s,a)$ gives the probability that the system will transition to state $s'$ after action $a$ is selected in state $s$. Here we assume a discrete time setting, where continuous time systems can be handled by discretizing at an appropriate time scale. Note that $P$ defines a first-order dynamic system in the sense that the distribution over next states only depends on the current state and the selected action, rather than also depending on previous states and actions. In an electrical network, $P$ describes network state transitions due to both electrical laws, the control actions, and random exogenous events. For example, an operator action of taking a transmission line out of service will result in transient state transitions, whereas a random (exogenous) weather event will also result in such transitions. The final MDP component, $R$, is a reward function that maps states to numeric rewards. In the context of power networks, $R(s)$ might measure the overall profit or quality-of-service associated with state $s$, noting that negative rewards can be used to model system costs.

Given an MDP, the objective is to derive a controller for the system that will accumulate the largest possible long term reward by selecting appropriate actions over time. More formally, a controller for an MDP is typically referred to as a policy $\pi$ which is a mapping from states to actions. Given a policy $\pi$ the MDP is controlled by selecting the actions dictated by the policy in states that arise according to the MDP dynamics. Thus, a policy $\pi$ defines a distribution over the possible state sequences, and in turn reward sequences, of the system over time. Intuitively, we would like to use a policy that produces sequences with large accumulated rewards, which is formalized via the concept of value functions. Each policy $\pi$ is associated with a value function $V^{\pi}$, which is a function from states to real-numbers, such that $V^{\pi}(s)$ measure the ``quality" of $\pi$ when started in state $s$. More formally, in this paper we will let $V^{\pi}(s)$ be the expected discounted infinite sum of rewards starting from state $s$ given by: 
\begin{equation}
\label{eqn2}
V^{\pi}(s) = E\left[\sum_{t=0}^{\infty} \beta^t R(s_t)\right]
\end{equation}
where $E[\cdot]$ is the expectation operator, $\beta \in (0,1)$ is a discount factor, and $s_t$ is a random variable denoting the state that results after starting in state $s$ and following actions dictated by $\pi$ for $t$ steps (so $s_0 = s$). The discount factor $\beta$ is included in the above expression to ensure that the infinite sum of rewards remains finite, which is accomplished by exponentially discounting future rewards at a rate of $\beta$. If a small value of $\beta$ is used, then only temporally near rewards will influence the value of a policy. Rather, in practice $\beta$ is often set to large values close to one to ensure that more distant rewards have significant impact on the value. 

A policy $\pi^*$ is said to be an optimal policy if in any state $s$ its value $V^{\pi}(s)$ is at least as good as that of any other policy. A fundamental theoretical result is that every MDP has an optimal policy, though there may be multiple optimal policies (all with the same value function). Given an MDP description, it is possible to compute an optimal policy using classic algorithms such as value iteration or policy iteration \cite{puterman2009markov}. Unfortunately, these algorithms are only applicable to relatively small MDPs (thousands to millions of states) and become impractical for enormous MDPs such as those that arise from large electrical networks. Further, it can be shown that computing optimal policies for such systems is formally computationally hard \cite{bylander1994}. Thus, approximate solution approaches are generally used in such situations. One such approximate solution is policy switching, which we now describe.

\subsection{Policy Switching}
\label{ps}
In many control applications, it is possible to obtain or create a set of diverse policies for which it is expected that at least one of the policies will perform well in any situation that may arise. For example, in the context of electrical networks, any existing RAS scheme (and all of their parameterizations) is a possible policy. The main challenge in exploiting such a policy set, in practice, is that it is difficult to define rules for determining which policy is the best for a given state of the system. Policy switching \cite{chang2004,king2013} is a control technique designed precisely for this situation. In particular, rather than attempt to compute an optimal policy, policy switching instead attempts to compute a policy that is at least as good in any state as any of the policies in a provided policy set.

Policy switching assumes the availability of both a policy set $\Pi=\{\pi_1,\pi_2,\ldots,\pi_n\}$ and an MDP simulator. The simulator allows for starting in any state $s$ and simulating any policy $\pi$ in the MDP, resulting in a state and reward sequence. Note that for stochastic MDPs, the simulator may produce different results each run.  Using such a simulator it is possible to compute an estimate $\hat{V}^{\pi}(s)$ of $\pi$'s value function $V^{\pi}(s)$ from any state $s$. In particular, this can be done by conducting some number of Monte-Carlo simulations of $\pi$ starting in $s$ for a fixed time horizon and averaging the total discounted rewards across the simulations. This estimate converges rapidly to the true value as the number of simulations and time horizon grows\footnote{Convergence is exponentially fast with respect to increasing the time horizon and the number of simulations.} \cite{kearns2002sparse}. The availability of a reasonably accurate MDP simulator is realistic in many domains, including power systems. There are continually improved research grade tools as well as mature industrial grade simulators that answer a variety of problems for power networks.

Given the ability to estimate values using the simulator, policy switching uses a very intuitive approach to selecting actions. In particular, when arriving at state $s$ in the actual system, policy switching computes the value estimate of each policy and then selects the action chosen by the highest valued policy. After selecting the action, the system then transitions to a new state and the policy switching process repeats. Note that since all policies are considered at each encountered state, it is possible to switch between different policies over time. We propose a correspondence between these policies and the actions dictated by a RAS scheme for a given period of time. Formally, the policy switching policy over policy set $\Pi$, denoted by $\Pi_{\mbox{ps}}$, is defined as follows:
\begin{equation}
\label{eqn3}
\Pi_{\mbox{ps}}(s) = \pi_{i^*}(s), \;\;\; i^* = \arg\max_i \hat{V}^{\pi_i}(s)
\end{equation}
where here $i^*$ is the index of the policy that that accumulates the maximum reward in state $s$ using our simulator to estimate policy values. The computation time to obtain the policy $\Pi_{\mbox{ps}}$ in state $s$ scales linearly with the number of policies being considered. The run time is typically dominated by the simulations required for value estimates. Importantly, the algorithm can be easily parallelized since all simulations can be run on independent processors. This allows for an approximate linear reduction in runtime in terms of the number of available processors. This is important if one wants to consider large policy sets while also meeting real-time constraints on action selection. 

It is not entirely obvious that it is safe in general to allow such free switching between any policy in a large set. However, this simple algorithm is able to provide the very strong guarantee that the value of the switching policy $\Pi_{\mbox{ps}}$ in any state is at least as good as that of the best policy in $\Pi$.
In particular, assuming that there is no approximation error in the value estimates $\hat{V}^{\pi_i}$ we get the following:
\vspace{10pt}
\begin{theorem}[Adapted from \cite{chang2004}]
\label{T1}
For any MDP, any policy set $\Pi$, and any state $s$ we have that $V^{\Pi_{\mbox{ps}}}(s) \geq \max_i V^{\pi_i}(s)$. 
\end{theorem}
\vspace{10pt}
Thus we can leverage the potential benefits of switching without a system penalty. The above guarantee assumes that the value estimates were perfect, which will not be the case in reality due to both inaccuracies in the simulator and Monte-Carlo sampling. However, the result has been extended \cite{king2013} to the case of approximate value estimates, where each value estimate $\hat{V}^{\pi}$ is guaranteed to be within $\epsilon$ of the true value. In that case, the above theorem is modified so that the value of the switching policy is no more than $2\epsilon$ worse in value than the best policy in the set.


\section{Methodology}
\label{methodology}

This section is meant to lay out the operation of the policy switching algorithm that has been implemented in PSS/E version 33. In order to perform efficient and organized experiments that interface with PSS/E in real time, we use a Python-based computational engine to complete the file organization, simulation controls, and post-processing. The PSS/E API is heavily used to automate data and file loading, network control actions (such as opening a transmission line), and data processing from PSS/E output.

First, we performed a dynamic study of the IEEE Case 39 benchmark system due to the complex transient characteristics that arise in power system contingency situations. As suggested in earlier RAS studies such as \cite{Paper013} and \cite{Paper002}, we chose to utilize a time domain, dynamic simulation. See Appendix \ref{appendix} for details about the specific dynamic parameters (i.e., generator machine models, governors, and exciters).

We have set up a simulation strategy over two RAS policies where we implement a Monte-Carlo process that parallely computes the best base policy for each dispatch time after a contingency has occurred. A contingency is said to have occurred, in the case of our experiments, when any single bus voltage falls outside of the acceptable operating range. The acceptable operating range, derived from both the US Western and Eastern Interconnection standards, is: $.9 < V_{pu} < 1.12$. Besides defining these values, we have also modified the line flow limits to ensure an $N-1$ secure network. Specifically, we added 10\% capacity to each branch's limit given any $N-1$ situation. 

Another important aspect of using policy switching as an effective power system protection algorithm is determining the effectiveness of each individual policy within each successive state (in this case, successive dispatch time intervals). As described in Sections \ref{mdp} and \ref{ps}, a reward value is associated with each successfully completed Monte-Carlo simulation of policy $\pi_i$ within the current state $s$. Thus given \textit{Theorem \ref{T1}} we simply need to define a universal reward method for the two RASs in our set of base policies. We have defined the reward value for each individual base policy as follows:
\begin{equation}
\label{eqn4}
 R(s_t) = \frac{L_t-L_{t-1}}{2 * L_{total}}  + \frac{B}{2} 
\end{equation} 
where $L_t$ and $L_{t-1}$ are the amounts of operational load at time $t$ and some arbitrary time horizon earlier $t-1$, respectively. $B$ is a boolean/binary variable that takes on the value $1$ or $0$ depending on whether the system was returned to a stable and acceptable operating point as a result of the policy.\footnote{In this context stable and acceptable is defined as all in-service bus voltages ($V$) satify the constraint: $0.9pu \leq V \leq 1.12pu$.} The constant $L_{\textnormal{total}}$ represents the total load demand in the case. In short, $R(s_t)$ is a reward for an individual policy in state $s$ at time $t$. It is defined as the weighted and normalized sum of the change in operational load over the time horizon $t-1$ to $t$, and whether or not the power system was returned to a nominal operating point. This, along with Eq.~\ref{eqn2} yields the value function ($V^{\pi}(s)$). This equation now spans one entire simulation ranging from time $0$ to $\tau$, therefore we take $\beta$ to be $1$ in this finite time horizon as follows:
\begin{equation}
\label{eqn5}
 V^{\pi}(s) = \sum_{t=0}^{\tau} R(s_t)
\end{equation}

\begin{figure}[!h]
\begin{center}
\includegraphics[scale=.41]{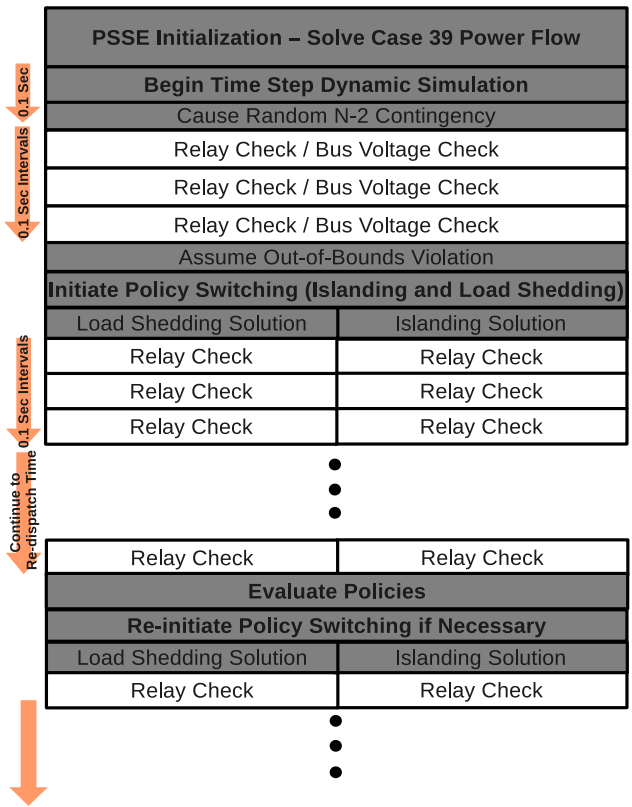}
\caption{\textit{This figure depicts an example formulation of the policy switching algorithm that has been implemented.}}
\label{fig2}
\end{center}
\end{figure}

Fig.~\ref{fig2} is a schematic visualization of the policy switching simulation technique. In summary, PSS/E is initialized with Case 39 information, a random N-2 contingency is simulated, and the policy switching algorithm works to recover the system with minimal load lost in each state.

Another key point to note about Fig.~\ref{fig2} is that the algorithm takes a continuous time, real-world application and breaks it into three discrete timescales.  The first, and lowest level (not pictured), is PSS/E's default time step of 1/120 seconds.  Next, is our defined relay operation time frame of 0.1 seconds. Notably, a sub-second timing is used since this is an acceptable range for which modern relays evaluate and take action in contingency situations. Third, is our policy switching re-dispatch (or intervention) time, which is derived from the settling time of the primary governor and exciter controls of the machines (5 seconds). See Appendix \ref{appendix} for specific values.

\section{Results}
\label{exp}

In this section we provide an in depth review of one of the $N-2$ contingency situations simulated on IEEE Case 39. First, we open the branches between Bus 19 and Bus 20, as well as between Bus 2 and Bus 25.  This subsequently causes an out-of-bounds violation for multiple buses throughout the network topology. Three experiments are then conducted to test the proposed policy switching approach.  In the first two experiments, we limit the policy space $\Pi$ to only one policy. The first policy is Islanding (I), and the second policy is load shedding (LS). In the third experiment, we increase the space of policies ($\Pi$) to include both islanding (I) and load shedding (LS). The value function for each experiment is calculated from the rewards of each policy in the individual experiments. These results are detailed in Table \ref{Table 1}.

\begin{table}[!h]
\caption{Value Functions Results}
\label{Table 1}
\centering
\begin{tabular}{c  p{2cm}  c  p{2cm}}
\toprule[1.5pt]
Experiment & Policies Available ($\Pi$) & $V^{\pi}(s)$ & Operational Load [MVA] \\
\cmidrule[1pt](rl){1-4}
1 & \multicolumn{1}{c}{I} & 0.296 & \multicolumn{1}{c}{3848} \\
2 & \multicolumn{1}{c}{LS} & 0.725 & \multicolumn{1}{c}{2922} \\
3 & \multicolumn{1}{c}{I \& LS} & 0.865 & \multicolumn{1}{c}{4749}\\
\bottomrule[1.5pt]
\end{tabular}
\end{table}

From the value function results in Table \ref{Table 1}, one can quickly identify the policy space $\Pi(I,LS)$ as the most suitable solution to the contingency. However, it is not entirely clear that the proposed algorithm does indeed select the best combination of RAS actions from those available. This is especially true in the case of Experiment 3 where there are multiple policies to choose from.  In order to show that the theory presented in Section \ref{back} can be intelligently utilized to guarantee a recovered and stable network, and that the methodology presented in Section \ref{methodology} is apt for this topology, we provide further data highlighting what occurs both in the switching algorithm and in the power network behind the scenes. These data are seen in Table \ref{Table 2}.

Table \ref{Table 2} ultimately shows that, via policy switching, one can achieve an efficient algorithm that  successfully implements multiple protection methods to meet and stabilize the unique transient characteristics of a specific outage. To create this cohesive enumeration of results we forced the algorithm to choose every possible combination of islanding and load shedding over our simulation time horizon. It is important to note that this is not the way in which the algorithm would operate in a real-world setting. By observing the changes in operational load after using different RAS policies, as well as by tracking the stability of the system (given by $B$), it becomes apparent that the policy switching algorithm chose the best order of policies to use at each consecutive dispatch time.  The shaded cells in Table \ref{Table 2} depict the path that our policy switching algorithm chose in Experiment 3.

\begin{table*}[!t]
\caption{Full Enumeration of the example N-2 Contingency -- Line 19 to 20 and Line 2 to 25}
\label{Table 2}
\centering
\begin{tabular}{ c  c  c  c  c  c  c  c  c }
\toprule[1.5pt]
Policy Available ($\Pi$): & Islanding (I) & Load Shedding (LS) & \multicolumn{6}{c}{Islanding and Load Shedding} \\
\cmidrule[1pt](rl){1-9}
\textbf{FIRST POLICY} & I & LS & I & I & I & LS & \cellcolor{red!50}LS & \cellcolor{red!50}LS \\
\cmidrule[1pt](rl){1-9}
\multicolumn{1}{l}{Operational Load ($L$) [MVA]} & 4759 & 5224 & 4759 & 4759 & 4759 & 5224 & 5224 & 5224 \\
\multicolumn{1}{l}{System Saved ($B$)} & NO & NO & NO & NO & NO & NO & NO & NO \\
\multicolumn{1}{l}{Reward$^\dagger$ ($V^\pi$)}& 0.366 & 0.401 & 0.366 & 0.366 & 0.366 & 0.401 & 0.401 & 0.401 \\
\multicolumn{1}{l}{\# Buses Lost} & 7 & 3 & 7 & 7 & 7 & 3 & 3 & 3 \\
\multicolumn{1}{l}{\# Generators Lost} & 2 & 2 & 2 & 2 & 2 & 2 & 2 & 2 \\
\multicolumn{1}{l}{\# Full Loads Lost} & 4 & 1 & 4 & 4 & 4 & 1 & 1 & 1 \\
\multicolumn{1}{l}{\# Transmission Lines Lost} & 13 & 6 & 13 & 13 & 13 & 6 & 6 & 6 \\

\cmidrule[1pt](rl){1-9}
\textbf{SECOND POLICY} & I & LS & I & LS & LS & LS & \cellcolor{red!50}I & \cellcolor{red!50}I \\
\cmidrule[1pt](rl){1-9}
\multicolumn{1}{l}{Operational Load ($L$) [MVA]} & 3848 & 2922 & 3848 & 4283 & 4283 & 2922 & 4749 & 4749 \\
\multicolumn{1}{l}{System Saved ($B$)} & NO & YES & NO & YES & YES & YES & YES & YES \\
\multicolumn{1}{l}{Reward$^\dagger$ ($V^\pi$)}& 0.296 & 0.725 & 0.296 & 0.829 & 0.829 & 0.725 & {\cellcolor{red!50}}0.865 & \cellcolor{red!50}0.865 \\
\multicolumn{1}{l}{\# Buses Lost} & 14 & 21 & 14 & 7 & 7 & 21 & 5 & 5 \\
\multicolumn{1}{l}{\# Generators Lost} & 3 & 5 & 3 & 2 & 2 & 5 & 2 & 2 \\
\multicolumn{1}{l}{\# Full Loads Lost} & 8 & 9 & 8 & 4 & 4 & 9 & 3 & 3 \\
\multicolumn{1}{l}{\# Transmission Lines Lost} & 21 & 27 & 21 & 13 & 13 & 27 & 9 & 9 \\

\cmidrule[1pt](rl){1-9}
\textbf{THIRD POLICY} & I\textsuperscript{*} & LS\textsuperscript{**} & LS & I\textsuperscript{**} & LS\textsuperscript{**} & I\textsuperscript{**} & LS\textsuperscript{**} & I\textsuperscript{**} \\
\cmidrule[1pt](rl){1-9}
\multicolumn{1}{l}{Operational Load ($L$) [MVA]} & - & - & 3463 & - & - & - & - & - \\
\multicolumn{1}{l}{System Saved ($B$)} & - & - & YES & - & - & - & - & - \\
\multicolumn{1}{l}{Reward$^\dagger$ ($V^\pi$)}& - & - & 0.766 & - & - & - & - & - \\
\multicolumn{1}{l}{\# Buses Lost} & - & - & 14 & - & - & - & - & - \\
\multicolumn{1}{l}{\# Generators Lost} & - & - & 3 & - & - & - & - & - \\
\multicolumn{1}{l}{\# Full Loads Lost} & - & - & 8 & - & - & - & - & - \\
\multicolumn{1}{l}{\# Transmission Lines Lost} & - & - & 21 & - & - & - & - & - \\

\bottomrule[1.5pt]
\multicolumn{9}{l}{\footnotesize{$\dagger$ Reward calculated based on $L_{total} = 6501$ [MVA] }}\\
\multicolumn{9}{l}{\footnotesize{\textsuperscript{*}No third islanding policy available -- Case too small}}\\
\multicolumn{9}{l}{\footnotesize{\textsuperscript{**}Not needed -- In-service buses within acceptable operating range}}
\end{tabular}
\end{table*}


\section{Conclusions and Future Work}
\label{conclude}

\subsection{Concluding Remarks}

The above work is meant to be a preliminary introduction to the fundamental idea of policy switching for power systems protection. In an effort to show the benefits and capability of policy switching, simple protection methods (islanding and load shedding) were implemented and tested on a small network topology using a well validated time-domain, dynamic simulator (Siemens PSS/E). 

We found that within any set of defined protection policies there is a unique combination of protection measures that optimally return the electric grid to an acceptable and stable state. Perhaps nonintuitively, this combination of protection measures is \textit{not necessarily} a single one-size-fits-all RAS. Rather, for the nearly infinite number of contingencies, both exogenous and endogenous, there will be a specific order of protection measures that must be implemented to optimally save the power network from off-nominal modes that have the potential to develop into cascading outages.

In conclusion, policy-switching provides an algorithmic means to determine which RAS policy should be implemented and avoids presupposing the individual solution to any specific contingency.  This is becoming increasingly important, especially as more complex constraints are introduced into grid operation such as variable renewable generation, large storage capability, increased grid congestion, and demand for high quality power in data centers, and fabrication laboratories. One question that still remains is: can we afford to simulate every combination of policies -- can we design this algorithm to effectively balance computational, economic, societal, and political expenses? 

\subsection{Future Work}

Finally, we list some important considerations that will be made in continuing this research and in improving upon the methodology explained in this paper.

\subsubsection{Algorithm Considerations}
\begin{itemize}
\item Increase the number and complexity of RAS policies in the policy set.
\item Define a complex and multi-faceted reward allocation scheme. 
\item Implement different percentages of variable renewable generation density.
\end{itemize}
\subsubsection{Simulator Considerations}
\begin{itemize}
\item Test algorithm on larger grid topologies (i.e., Poland, Ireland, and larger IEEE test cases).
\item Perform quantitative investigation of computational complexity.
\item Increase parallelization capabilities especially via supercomputing resources.
\end{itemize}

\section*{Acknowledgments}
The authors would like to acknowledge partial support from Oregon BEST/Bonneville Power Administration NW Energy Experience Scholarship, as well as partial support from NSF grant IIS-0964457.

\bibliographystyle{IEEEtran}
\IEEEtriggeratref{15} 
\bibliography{PSCC14Database}

\appendix[Dynamic Model Values]
\label{appendix}
\vspace{.15in}
\begin{center}
\begin{tabular}{ c  c  c  c  c }
\toprule[1.5pt]
\multicolumn{5}{c}{'GENROU' Model -- Machine Parameters} \\
\cmidrule[1pt](rl){1-5}
\multicolumn{2}{c}{Machine Specific} & & \multicolumn{2}{c}{General} \\
\cmidrule[1pt](rl){1-5}
BUS & H &  & Parameter & Value \\
\cmidrule[1pt](rl){1-5}
30 & 42   &  & T'$_{do}$       & 5 \\
31 & 30.3 &  & T"$_{do}$       & 0.05 \\
32 & 35.8 &  & T'$_{qo}$       & 1.5 \\
33 & 28.6 &  & T"$_{qo}$       & 0.05 \\
34 & 26   &  & X$_{d}$         & 0.6 \\
35 & 34.8 &  & X$_{q}$         & 0.55 \\
36 & 26.4 &  & X'$_{d}$        & 0.075 \\
37 & 24.3 &  & X'$_{q}$        & 0.1125 \\
38 & 34.5 &  & X"$_{d}=$X"$_q$ & 0.05 \\
39 & 50   &  & X$_l$           & 0.001 \\
-  & -    &  & S(1.0)          & 0.11 \\
-  & -    &  & S(1.2)          & 0.48 \\
-  & -    &  & D               & 1 \\
\toprule[1.5pt]
\multicolumn{5}{c}{Governor and Exciter Models\textsuperscript{$\dagger$}} \\
\cmidrule[1pt](rl){1-5}
\multicolumn{2}{c}{'IEEEG2'} & & \multicolumn{2}{c}{'IEEET1'} \\
\cmidrule[1pt](rl){1-5}
Parameter & Value & & Parameter & Value \\
\cmidrule[1pt](rl){1-5}
K       & 20   &  & T$_{R}$        & 0.1 \\
T$_1$   & 50   &  & K$_{A}$        & 100 \\
T$_2$   & 9.99 &  & T$_{A}$        & 0.02\\
T$_3$   & 1    &  & V$_{RMAX}$     & * \\
P$_MAX$ & 20   &  & V$_{RMIN}$     & 0 \\
P$_MIN$ & 0    &  & K$_{E}$        & * \\
T$_4$   & 1    &  & T$_{E}$        & 0.1 \\
-       & -    &  & K$_{F}$        & 0.03 \\
-       & -    &  & T$_{F}$        & 1 \\
-       & -    &  & Switch         & 0 \\
-       & -    &  & E$_{1}$        & 2.9\\
-       & -    &  & S$_{E}$E$_{1}$ & 0.5\\
-       & -    &  & E$_{2}$        & 3.9\\
-       & -    &  & S$_{E}$E$_{2}$ & 0.86\\
\bottomrule[1.5pt]
\multicolumn{5}{l}{\footnotesize{\textsuperscript{$\dagger$}Identical governors and exciters used for all machines.}}\\
\multicolumn{5}{l}{\footnotesize{\textsuperscript{*} Denotes PSS/E automatically determines this value.}}

\end{tabular}
\end{center}

\end{document}